\documentclass[epjCONF]{svjour}
\usepackage{graphics}
\usepackage[varg]{txfonts} % Times fonts
\usepackage[latin1]{inputenc}
\usepackage{xspace}

\def\superb{Super$B$\xspace}

\usepackage{relsize}
\def\babar{\mbox{\slshape B\kern-0.1em{\smaller A}\kern-0.1em
    B\kern-0.1em{\smaller A\kern-0.2em R}}\xspace}

\session-title{he complementarity of \superb with the LHC}
\begin{document}
\title{\boldmath The complementarity of \superb with the LHC}
\author{Adrian Bevan\inst{1}\fnmsep\thanks{\email{a.j.bevan@qmul.ac.uk}}}
\institute{School of Physics and Astronomy, Queen Mary University of London, London, E1 4NS, UK.}
\abstract{
The complementarity between results anticipated from \superb with those from
the LHC experiments is discussed here. \superb can contribute to searches
for new physics using indirect constraints via precision tests of the 
standard model.  In addition to the indirect constraints, there are a number
of direct searches that can be performed at low energy.  There is a well 
motivated programme of measurements to make at \superb, the results of which
will add to our understanding of possible scenarios of physics beyond the standard 
model.
} %end of abstract
\maketitle

\section{Introduction}
\label{sec:bevan:intro}
The \superb project is described in detail in Refs~\cite{acceleratorwp,detectorwp,physicswp,sbimpact}.  
This is a high luminosity $e^+e^-$ experiment designed to study both quark and charged lepton 
flavour transitions with vast numbers of $B_{u,d,s}$, $D$ and $\tau$ decays, as well as perform precision tests
of the standard model of particle physics (SM).  A few of the measurements 
possible at \superb can be performed at the LHC, however the real strength
of the physics programme rests in the observables that are unique to these 
facilities.  These proceedings concentrate on some observables accessible 
to \superb and in particular measurements of these complement
the knowledge being attained at the LHC.  This discussion is broken down
into issues pertaining to the new physics (NP) energy scale, followed by forbidden
and rare processes, CP violation and mixing, and precision electroweak 
constraints.  A more detailed discussion of the 
\superb physics programme can be found in Refs~\cite{physicswp,sbimpact}. 
It is anticipated that \superb will be taking data by 2017.

As discussed in section~\ref{sec:bevan:energyscale}, one may learn
something about the energy scale of any underlying NP from such 
measurements, should this not already be ascertained from LHC results 
by the time data taking commences.  If the NP energy scale is defined by results
from the LHC before the Super Flavor Factories take data,
one can start to constrain couplings within the underlying theory.
Otherwise deviations from the SM may give some indication of an upper 
bound on the NP scale and hence the ideal integrated luminosity required to observe NP at the SLHC.
Forbidden and rare decays within the SM constitute a set of 
powerful quantum interferometers for NP searches.
In the case of forbidden processes (section~\ref{sec:bevan:forbidden})
one tests for null results, while in the case of rare decays 
(section~\ref{sec:bevan:rare}) one searches for deviations from 
SM expectations resulting from interference effects.  Studies of such decays 
are cleaner in an $e^+e^-$ environment than a hadronic one, and in many cases
provide complementary information to direct or indirect searches
at the LHC.  An advantage of studying decays in a clean environment
is that in addition to all charged hadronic final states, that are often
subject to non-trivial model uncertainties, one can measure
processes that are theoretically clean.  Such probes for
NP provide a robust set of redundant measurements that can be 
used to disentangle the nature of any underlying NP that may be manifest
(see the interplay discussions in Refs~\cite{physicswp,sbimpact}).
While mixing and CP violation phenomena have been extensively studied 
in neutral kaon and $B$ meson systems at previous generations of 
experiments, there has been relatively little work done in 
the charm sector.  Similarly only a few CP violation measurements have been made in
$\tau$ decays. \superb will be able to probe these effects in $B$, $D$ and $\tau$ decays
as discussed in section~\ref{sec:bevan:cpvmixing}.
Finally \superb's potential for precision electroweak
physics is highlighted.

\section{The new physics energy scale}
\label{sec:bevan:energyscale}

Naturalness arguments have led theorists to postulate that new heavy 
particles must exist at the electroweak symmetry breaking scale, with masses of $\leq 1 TeV$.
Recent results from the LHC have failed to uncover any sign of such new particles,
and it is expected that searches will reach the TeV scale soon.  
One should recall that the interpretation of these 
direct searches for NP are done in a model dependent way, for example
using the so-called constrained minimal SUSY scenario (CMSSM) or some other 
simplified model.
Nonetheless experimental results from the LHC place lower
bounds on the masses of postulated new particles.  If one considers, for example, 
generic MSSM in the mass-insertion hypothesis~\cite{Hall:1985dx,Ciuchini:2002uv}, there are well in excess of
100 parameters that need to be considered as opposed to just five used for CMSSM
searches.  Most of these parameters are complex couplings in the flavour sector
that are often ignored in phenomenological studies.
The reason why phenomenologists do not consider generic studies of 
the full parameter space is the result of practical issues, i.e. numerical considerations
and resources. The well known
curse of dimensionality~\cite{bellman} is an issue affecting analysis where the number of samples in a given dimension 
$n$ ($\sim 1000$ or larger) raised to the power of the number of dimensions $m$ ($>100)$.  
Hence the total number of samples required for a complete study is $n^m > 10^{300}$
which clearly places a practical limit on phenomenological analysis.

We can consider the implications of indirect constraints on NP attainable 
through the study of low energy effects, where for example historically the large 
value of $\Delta m_d$ measured by the ARGUS experiment contradicted 
popular belief that the top quark was light (a few tens of GeV), and the result of that story
was the discovery of a heavy top quark with mass $\sim 170$ GeV, just where indirect
constraints had predicted it would be.  Today 
there are no large deviations from loop dominated flavour changing 
processes that point to a TeV scale NP and this needs to be understood.
Either there is NP at a TeV, in which case we need to discover this 
at the LHC and subsequently understand why flavour has not 
revealed its presence indirectly as was often the case in the past, 
or there is no NP at that scale and we need to revert to 
indirect searches using lower energy probes at various experiments.

An example of this is illustrated in the following:
Complex couplings present in MSSM in the mass-insertion hypothesis can be
constrained using a number of rare decays, for example inclusive measurements
of $b \to s \gamma$ and $b \to s \ell\ell$.  While existing measurements
do not place significant constraints on couplings, with data from \superb one 
will be able to measure the magnitude and phase of some of these couplings,
and simultaneously place an upper bound on the mass of new particles.
%, and one can also infer 
%an upper bound on the mass of new particles (e.g. a gluino).  
Hence if the LHC discovers a $\tilde{g}$ in the near
future, the inclusive $B$ decays mentioned here can be used to determine
a coupling constant of the model.  If however there is no imminent 
discovery from CERN, then results from \superb could also be used to infer an upper limit 
on the mass scale to complement direct bounds from the LHC.  In turn
this information can be used to determine how much data one would 
require from the SLHC in order to make a direct discovery of a $\tilde{g}$
in this model.  This is a single example of the so-called
interplay problem linking results from particle physics and 
cosmological studies in the context of elucidating the structure of some 
higher theory.  Further discussion on this issue can be found in
~\cite{physicswp,sbimpact} and references therein.

\section{Forbidden processes}
\label{sec:bevan:forbidden}

While it is well known that quarks and neutrinos mix, and that this 
phenomenon of weak interactions has profound impact on the manifest 
nature of the universe, as yet there is no direct evidence to support the 
hypothesis that charged leptons may change type.  However the fact that 
neutrinos mix means that lepton number conservation is violated in nature.  
A natural consequence of
this is to expect that one may ultimately learn that lepton number 
is also violated in charged decays, this is usually referred to as
lepton flavour violation (LFV).  Given three generations of charged lepton,
there are three couplings associated with charged LFV relating
to (i) $\mu \to e$ (ii) $\tau \to \mu$, and (iii) $\tau \to e$ 
transition probabilities.  This area of physics is highly model 
dependent, and so it is imperative that one searches for processes
related to all three of these couplings in order to determine 
both the magnitude and hierarchy of any possible effects.  The MEG
experiment at PSI is searching for transitions of the first type
via $\mu \to e \gamma$~\cite{Adam:2011ch} and is expected to reach an ultimate 
sensitivity of $\sim 10^{-13}$.  The COMET and Mu2E experiments
aim to study $\mu \to e$ transitions via conversion in material.
\superb will be able to search for a wide range of complementary 
LFV signatures to constrain
couplings related to $\tau \to \mu$ and $\tau \to e$ transitions.
A unique feature of \superb is a polarised electron beam, which in turn
allows one to reconstruct the polarisation of the decaying taus.
Using this information one can suppress SM background 
in searches for $\tau\to \mu \gamma$ decays, as well as
probe the chiral structure of any NP encountered.

In many models one transition will dominate over the others,
highlighting the relevance of searching for both $\mu$ and $\tau$ charged
LFV processes.  Understanding of the full set of transitions is ultimately 
required in order to establish a theory of charged LFV.
The manifest phenomena can also be correlated with 
the neutrino sector, and with quark flavour changing processes e.g. see Ref.~\cite{physicswp},
and references therein.
In specific models the level of LFV manifest
in the $\tau$ sector can be related to CP violation and mixing in
the $B_s$ sector, highlighting the 
importance of a global (quark and lepton) flavour analysis.

\section{Rare decays}
\label{sec:bevan:rare}

There are a number of interesting rare decays of $B$ and
$D$ mesons that may be affected by NP and can be studied at \superb.  
Rare charm decays often (but not always) have theoretical uncertainties arising from 
long distance contributions that may make interpretation of the data challenging.
Nonetheless they can be used to probe dynamics of various NP scenarios.  The
golden rare decay channels for \superb include inclusive $b\to s \gamma$ and
$b\to s \ell \ell$ decays as discussed in Section~\ref{sec:bevan:energyscale},
which can only be measured in a clean $e^+e^-$ environment.  In addition
to these one can constrain the charged Higgs mass using a combination of
$B\to \tau \nu$ and $B\to \mu \nu$.  This will complement the direct searches and 
increase the energy range probed to the multi-TeV level ahead of the SLHC.
One can study the Z-penguin operator sector of the SM using $B\to K\nu\overline{\nu}$
decays, and in doing so, cleanly search for NP that may manifest itself here
(for example $Z^\prime$ and new heavy scalar particles).

The LHCb experiment will be able to study $B\to K^{(*)} \mu^+\mu^-$, and is expected
to accumulate of the order of 8,000 events in $5fb^{-1}$ of data. \superb is expected
to accumulate between 10,000 and 15,000 events in both the di-muon and di-electron final 
states (as well as making measurements of the corresponding inclusive modes).  The
combination of all four of these measurements is required in order to disentangle
the full set of possible NP models that can affect these processes.  Furthermore,
by measuring all final states, one has the opportunity to understand if NP is 
present, or if one is witnessing a statistical fluctuation.

One can also study $B_s$ rare decays using data from $\Upsilon(5S)$ running, such as $B_s\to \gamma\gamma$, which can be enhanced
by SUSY and 2HDM.  The correlation between observed enhancements in $B_s$ rare decays,
with the corresponding $B_d$ processes such as $b\to (s,d)\gamma$ can be used to distinguish between models.
Detailed discussions of the rare decay programme at \superb can be found in 
Refs~\cite{physicswp,sbimpact}.

\section{CP violation and mixing}
\label{sec:bevan:cpvmixing}

\superb will be able to study CP violation in $B$, $D$ and $\tau$ decays.
Expectations for the precision attainable for both time-dependent and 
time-integrated measurements in $B^0$ and $B^\pm$ decays at \superb
are well understood, based on a decade of groundwork laid by 
the \babar and Belle experiments, and can be found in Ref.~\cite{physicswp}.
One may think that the previous decade of measurements would have negated
the interest in this area, however it has recently been pointed out
that the golden mode ($J\psi K_S^0$) measurement of $\sin 2\beta$
is $\sim 3\sigma$ from the SM value~\cite{Lunghi:2007ak}.  Not only that, but one can 
search for NP via $b\to c$ tree, as well as $b\to d$ 
and $b\to s$ loop transitions.

Knowledge of charm meson mixing is relatively recent, and as a result
the next generation of experiments is expected to make significant progress
not only in studying mixing, but also searching for possible CP violation
effects.  The channel that dominates our knowledge of mixing is $D\to K_S h^+h^-$, which 
requires input from measurements at charm threshold, $\psi(3770)$.  This vital input is
a map of the strong phase difference as a function of position in
the $K_S h^+h^-$ Dalitz plot.  This strong phase map is also an input into
the programme to measure the unitarity triangle angle $\gamma$.  Recently
it was remarked that one can perform time-dependent CP asymmetry measurements
in charm decays in analogy with the approach taken by the B factories, and that such 
time-dependent measurements may provide a new way to directly measure 
the mixing phase~\cite{Bevan:2011up}.
Following this presentation LHCb showed results on a non-zero asymmetry 
difference between $D\to \pi^+\pi^-$ and $D\to K^+K^-$ decays~\cite{Aaij:2011in}.  
That measurement does not use all of the available information and it is hoped that LHCb will
adopt the methodology proposed for \superb in the future in order to 
disentangle direct and indirect CP asymmetries.  Understanding
if one has observed NP or not will require inputs from a number of different
channels, including all neutral final states, motivating the need for results from \superb.

Searches for signs of CP violation in tau lepton decays have been largely neglected
thus far.
A recent measurement from \babar shows a $3\sigma$ deviation from SM expectations 
in $\tau\to K_S\pi^0\nu$ decays~\cite{BABAR:2011aa}, where as the corresponding Belle result (using
a different approach) is compatible with expectations~\cite{Bischofberger:2011pw}.  \superb
will be able to perform precision tests of CP violation in the $\tau$ sector 
to compliment work done with $B$ and $D$ mesons.

\section{Precision electroweak constraints}
\label{sec:bevan:precisionew}

One of the inputs to precision electroweak fits used to predict the 
SM Higgs mass is the measurement of $\sin^2\theta_W$ from LEP/SLC.
The $e^+e^- \to b\overline{b}$ contribution to this result is affected
by hadronization uncertainties that limit the power of the experimental
measurement at the $Z$ pole.  It is possible to measure $\sin^2\theta_W$
using the full set of $e^+e^-$ to di-fermion transitions at the $\Upsilon(4S)$
energy where one expects to achieve the same precision on $\sin^2\theta_W$
as obtained for $e^+e^- \to b\overline{b}$ with the LEP/SLC combination.
Once the Higgs mass has been measured, $\sin^2\theta_W$ becomes a sensitive 
probe for NP~\cite{mike}.

\section{Summary}
\label{sec:bevan:summary}

The broad physics programme of \superb will be able to provide a number of 
constraints on NP to complement the direct searches and other measurements being
performed at the LHC.  If the LHC doesn't find direct evidence of NP before
\superb has accumulated a significant fraction of its data set, then it should
be possible to infer upper bounds on the NP energy scale.  With regard to flavour
measurements, \superb and LHCb complement each other as one experiment excels
at measurements of inclusive final states, states with many neutral 
particles ($\pi^0$, $\gamma$, $\nu$), and states with electrons in that would
otherwise be difficult to trigger on in a hadronic environment. Whereas the other 
provides precision constraints on final states with sufficient charged tracks 
to efficiently trigger on the event.  Combining results of results from \superb
with other flavour experiments will enable 
theorists to understand the structure of the NP Lagrangian.

%\begin{figure}
%% Use the relevant command for your figure-insertion program
%% to insert the figure file.
%% For example, with the option graphics use
%%\resizebox{0.75\columnwidth}{!}{%
%%  \includegraphics{fig1.eps} }
%\caption{Please write your figure caption here.}
%\end{figure}

%\begin{table}
%\caption{Please write your table caption here.}
%\begin{tabular}{lll}
%\hline\noalign{\smallskip}
%\end{tabular}
%\end{table}

\end{document}